\documentclass[aps,prc,preprint,nofootinbib]{revtex4-1}

\usepackage{sidecap}

\usepackage{ulem}
\usepackage{epsfig}
\usepackage{amsmath,amssymb,amsthm,mathtools}
\usepackage{graphicx}

\usepackage{xcolor}
\definecolor{ao}{rgb}{0.0, 0.5, 0.0}

\usepackage[hidelinks]{hyperref}

\DeclareMathAlphabet{\mathpzc}{OT1}{pzc}{m}{it}

\definecolor{ao}{rgb}{0.0, 0.5, 0.0}

\begin{document}



\newcommand{\vAi}{{\cal A}_{i_1\cdots i_n}}
\newcommand{\vAim}{{\cal A}_{i_1\cdots i_{n-1}}}
\newcommand{\vAbi}{\bar{\cal A}^{i_1\cdots i_n}}
\newcommand{\vAbim}{\bar{\cal A}^{i_1\cdots i_{n-1}}}
\newcommand{\htS}{\hat{S}}
\newcommand{\htR}{\hat{R}}
\newcommand{\htB}{\hat{B}}
\newcommand{\htD}{\hat{D}}
\newcommand{\htV}{\hat{V}}
\newcommand{\cT}{{\cal T}}
\newcommand{\cM}{{\cal M}}
\newcommand{\cMs}{{\cal M}^*}
\newcommand{\vk}{\vec{\mathbf{k}}}
\newcommand{\bk}{\bm{k}}
\newcommand{\kt}{\bm{k}_\perp}
\newcommand{\kp}{k_\perp}
\newcommand{\km}{k_\mathrm{max}}
\newcommand{\vl}{\vec{\mathbf{l}}}
\newcommand{\bl}{\bm{l}}
\newcommand{\bK}{\bm{K}}
\newcommand{\bb}{\bm{b}}
\newcommand{\qm}{q_\mathrm{max}}
\newcommand{\vp}{\vec{\mathbf{p}}}
\newcommand{\bp}{\bm{p}}
\newcommand{\vq}{\vec{\mathbf{q}}}
\newcommand{\bq}{\bm{q}}
\newcommand{\qt}{\bm{q}_\perp}
\newcommand{\qp}{q_\perp}
\newcommand{\bQ}{\bm{Q}}
\newcommand{\vx}{\vec{\mathbf{x}}}
\newcommand{\bx}{\bm{x}}
\newcommand{\tr}{{{\rm Tr\,}}}
\newcommand{\bc}{\textcolor{blue}}
\newcommand{\rc}{\textcolor{red}}

\newcommand{\beq}{\begin{equation}}
\newcommand{\eeq}[1]{\label{#1} \end{equation}}
\newcommand{\ee}{\end{equation}}
\newcommand{\bea}{\begin{eqnarray}}
\newcommand{\eea}{\end{eqnarray}}
\newcommand{\beqar}{\begin{eqnarray}}
\newcommand{\eeqar}[1]{\label{#1}\end{eqnarray}}
\newcommand{\nn}{\nonumber}
\newcommand{\tcr}{\textcolor{red}}
\newcommand{\dif}{\mathrm{d}}
\newcommand{\RAA}{$R_{AA}$}

\newcommand{\half}{{\textstyle\frac{1}{2}}}
\newcommand{\ben}{\begin{enumerate}}
\newcommand{\een}{\end{enumerate}}
\newcommand{\bit}{\begin{itemize}}
\newcommand{\eit}{\end{itemize}}
\newcommand{\ec}{\end{center}}
\newcommand{\bra}[1]{\langle {#1}|}
\newcommand{\ket}[1]{|{#1}\rangle}
\newcommand{\norm}[2]{\langle{#1}|{#2}\rangle}
\newcommand{\brac}[3]{\langle{#1}|{#2}|{#3}\rangle}
\newcommand{\hilb}{{\cal H}}
\newcommand{\pleft}{\stackrel{\leftarrow}{\partial}}
\newcommand{\pright}{\stackrel{\rightarrow}{\partial}}

\title{Signatures of nuclear structure and pre-equilibrium dynamics in O+O and Ne+Ne collisions at $\sqrt{s_{\rm{NN}}}$  = 5.36 TeV}

\author{Bithika Karmakar}
\email{bithika.karmakar@cvut.cz}
\author{Iurii Karpenko}
\email{karpeiur@cvut.cz}
\affiliation{Faculty of Nuclear Sciences and Physical Engineering,Czech Technical University in Prague, B\v{r}ehov\'a 7, 11519 Prague 1, Czech Republic}

\begin{abstract}
We study O+O and Ne+Ne collisions at $\sqrt{s_{\rm NN}}=5.36$ TeV within T$_{\text{R}}$ENTo+VISHNew+UrQMD framework, using Woods-Saxon, NLEFT and PGCM descriptions of ${}^{16}$O and ${}^{20}$Ne. The nuclear configurations are more clearly distinguished in the flow coefficients of Ne+Ne than of O+O collisions. The bowling-pin shape of ${}^{20}$Ne enhances $v_2\{2\}$(Ne+Ne/O+O) and suppresses $\rho(v_2\{2\}^2, \langle p_T \rangle)$ in central collisions, while the octupole contributions of both nuclei largely cancel in $v_3\{2\}$(Ne+Ne/O+O). Although $\langle p_T \rangle$ and the individual flow coefficients depend strongly on the pre-equilibrium dynamics, the ratio $v_2\{2\}$(Ne+Ne/O+O) ratio and the difference $\rho_2$ (Ne+Ne) - $\rho_2$ (O+O) are almost unaffected by it, establishing them as reasonably stable probes of nuclear structure in light-ion collisions. In contrast, the ratio $v_3\{2\}$(Ne+Ne/O+O) shows mild sensitivity to the medium effects, making it a less reliable probe of nuclear structure. 
\end{abstract}

\maketitle
\section{Introduction}

In Au+Au collisions at the Relativistic Heavy Ion Collider (RHIC) and in Pb+Pb collisions at the Large Hadron Collider (LHC), the measured anisotropic-flow coefficients and their event-by-event fluctuations are quantitatively reproduced by relativistic viscous hydrodynamical models. This agreement is an indirect evidence that the quark-gluon plasma (QGP) formed in these collisions behaves as a strongly coupled, nearly perfect fluid, with a specific shear viscosity which is constrained \cite{Bernhard:2016tnd,JETSCAPE:2020mzn,Nijs:2020ors} to be higher but close to the conjectured lower bound $\eta/s = 1/4\pi$~\cite{Kovtun:2004de}.

Signatures of collectivity have also been observed in so-called small collision systems. These include high-multiplicity pp and p-Pb collisions at the LHC and p+Au, d+Au and $^{3}$He-Au collisions at RHIC. The signatures comprise long-range near-side ("ridge") correlations and non-vanishing multiparticle cumulants~\cite{CMS:2010ifv,PHENIX:2018lia,CMS:2015yux,ALICE:2019zfl}. Their interpretation, however, remains under debate. Hydrodynamic and transport models capture the qualitative features of the data. Quantitative conclusions are nevertheless hampered by the poorly constrained initial-state geometry, in particular the spatial and subnucleonic structure of proton~\cite{Nagle:2018nvi,Grosse-Oetringhaus:2024bwr,Mantysaari:2022ffw}. Also, the Knudsen number is estimated to be not small for most of the evolution in small systems \cite{Niemi:2014wta}, which challenges the application of fluid dynamic approximation. 

Collisions of light ions, such as $^{16}$O+$^{16}$O and $^{20}$Ne+$^{20}$Ne, bridge the gap between these two regimes. They allow the system-size dependence of collective observables to be studied for nucleus--nucleus collisions with a well-defined overlap region. In such collisions the initial-state geometry is governed predominantly by the distribution of nucleons rather than by the poorly known substructure of the proton~\cite{Schenke:2021mxx}. 

A further motivation is to explore the nontrivial intrinsic structure of the colliding nuclei. The $J^{\pi}=0^{+}$ ground states of $^{16}$O and $^{20}$Ne are spherically symmetric in the laboratory frame \cite{Dobaczewski:2025rdi}. Nevertheless, \textit{ab initio} calculations within nuclear lattice effective field theory (NLEFT) and the projected generator coordinate method (PGCM) predict pronounced $\alpha$-cluster correlations in the intrinsic frame. These correspond to an (irregular) tetrahedral arrangement of four $\alpha$ clusters in $^{16}$O and an elongated, bowling-pin-like $\alpha$+$^{16}$O configuration in $^{20}$Ne~\cite{Lee:2025req,Frosini:2021sxj,Giacalone:2024luz}. The passage time of the colliding nuclei ($\sim 10^{-26}$~s) is much shorter than their intrinsic rotational time scales ($10^{-21}$--$10^{-20}$~s). Each collision therefore probes a single, randomly oriented intrinsic configuration, which is imprinted on the geometry of the initial overlap region~\cite{ALICE:2025luc}. The larger quadrupole deformation of $^{20}$Ne is thus expected to enhance elliptic flow in central Ne+Ne collisions relative to O+O collisions. The tetrahedral $^{16}$O, in turn, carries a larger octupole deformation~\cite{Giacalone:2024luz}.

The first LHC measurements in light-ion collisions, recorded in July 2025 at $\sqrt{s_{\rm NN}} = 5.36$~TeV, revealed sizeable elliptic ($v_2$) and triangular ($v_3$) flow in both systems~\cite{ALICE:2025luc,CMS:2025tga}. The ALICE Collaboration compared its data with hydrodynamic predictions obtained with the \textsc{Trajectum} framework. In these calculations the model parameters were not tuned to light-ion data; instead they were sampled from a Bayesian posterior constrained by Pb+Pb measurements~\cite{Nijs:2021clz}, and the nuclear configurations were taken from NLEFT and PGCM frameworks. With NLEFT input, these predictions reproduce both the magnitude and the centrality dependence of $v_2\{2\}$, $v_3\{2\}$ and $v_2\{4\}$ up to 50\% centrality~\cite{ALICE:2025luc}.

The imprint of nuclear structure is, by comparison, a small effect, confined mainly to the most central collisions. The measured ratio $v_2\{2\}_{\rm Ne+Ne}/v_2\{2\}_{\rm O+O}$ reaches about 1.08 in ultracentral events and decreases to about 1.05 at 10\% centrality~\cite{ALICE:2025luc}. This ratio is overestimated by the \textsc{Trajectum} predictions, which give $1.174$ (NLEFT) and $1.139$ (PGCM) in the 0--1\% centrality class~\cite{Giacalone:2024luz}. It is, however, better described by IP-Glasma+JIMWLK+MUSIC+UrQMD calculations. The difference has been attributed primarily to the description of the initial state, in particular to differences in subnucleonic width settings between the two models~\cite{ALICE:2025luc}.

Despite this progress, the fluid-dynamic description of light-ion collisions relies to a large extent on a very limited number of frameworks. To our knowledge, the most systematic multi-observable calculations, including identified-hadron spectra and mean transverse momenta, have been performed with \textsc{Trajectum}~\cite{Nijs:2021clz,Giacalone:2024luz}. Moreover, Ne+Ne/O+O ratios are designed to cancel final-state effects and model uncertainties~\cite{Giacalone:2024luz,ALICE:2025luc}. Therefore, the sensitivity of the absolute observables to the individual model ingredients has therefore received comparatively little attention.


The aims of this paper are threefold. First, we provide a complementary post-diction from a different hydro-based model for flow observables in O+O and Ne+Ne collisions. For this we use a framework that is physics-wise comparable to Trajectum but differs in its numerical implementation. More importantly, its parameter setup is derived from a different Bayesian analysis using the same Pb+Pb and p+Pb data. Second, we disentangle how the nuclear structure assumptions and pre-equilibrium free streaming drive the flow observables individually. We also quantify the isolated effect of nuclear structure on the absolute observables, in addition to their Ne+Ne/O+O ratios, and show the ``baseline'' results with plain nuclear substructure (no clustering). Third, this paper documents in detail the model setup employed in the femtoscopic study of Ref.~\cite{Kincses:2026guk}.

The initial conditions and hydrodynamical setup are described in Sec.~\ref{sec_setup}. The results are presented and discussed in Sec.~\ref{sec_result}, and the main findings are summarised in Sec.~\ref{sec_summary}.

\section{Initial condition and hydrodynamic setup}
\label{sec_setup}
The initial entropy density profiles for ${}^{16}\text{O}$ and ${}^{20}\text{Ne}$ have been generated using T$_{\text{R}}$ENTo initial condition model~\cite{Moreland:2014oya}. Three different nuclear configurations have been used: i) the set of nucleon positions for each nuclei is chosen by sampling an uncorrelated Woods-Saxon (WS) distribution with exclusion distance $d=0.5$ fm; ii) the clustered nuclear configurations found within NLEFT (Nuclear Lattice Effective Field Theory) simulations; and iii) the $\it{ab\ initio}$ PGCM (Projected Generator Coordinate Method) formalism~\cite{Giacalone:2024luz}. For each case, $10^5$ minimum bias events are generated using  T$_{\text{R}}$ENTo, where each clustered nuclear configuration is randomly rotated. The T$_{\text{R}}$ENTo normalisation parameter has been tuned to get agreement with the charged particle multiplicities at midrapidity in $0-10\%$, $10-20\%$, $20-30\%$ and $30-40\%$ centrality classes~\cite{Modak:2026ymk}. For the other parameters, we use the maximum a posteriori (MAP) values found in  Ref.~\cite{Moreland:2018gsh}, which was conducted using the same modelling framework. Pre-equilibrium evolution is modelled by free-streaming until $\tau=0.48$ fm~\cite{Broniowski:2008qk}. Thereafter, the system is forcefully  equilibrated (sudden
equilibration) according to Landau matching condition i.e., the free-streaming energy-momentum tensor is replaced  by the form
from perfect hydrodynamics~\cite{Broniowski:2008qk}. The sudden equilibration causes the development of anisotropic collective flow velocity in the system. 
To study the effect of pre-equlibrium free-streaming on the flow coefficients and mean transverse momentum in light-ion collisions, we also consider two additional scenarios: one with a free-streaming velocity of $v_{fs}=0.7$ and another without pre-equilibrium free-streaming.

Event-by-event hydrodynamical evolution of the medium is performed using VISHNew~\cite{Song:2007ux, Song:2008si, Shen:2014vra} where the dissipative fluid-dynamical equations are solved assuming longitudinal boost invariance:
\begin{equation}
  \partial_\mu T^{\mu\nu} = \partial_\mu\left(e\, u^\mu u^\nu - (p + \Pi)\,\Delta^{\mu\nu} + \pi^{\mu\nu}\right)=0.
\end{equation}
Here, $e$ and $p$ are the local energy density and thermal equilibrium pressure, $u^{\mu}$ is the four velocity of the energy flow, $\Pi$ is the bulk pressure and $\pi^{\mu\nu}$ is the traceless shear stress tensor. The shear stress and bulk pressure are evolved according to the Israel-Stewart~\cite{Israel:1979wp} type equations:
\begin{eqnarray}
\Delta^{\mu\alpha} \Delta^{\nu \beta} D \pi_{\alpha \beta} &=& -\frac{1}{\tau_\pi} (\pi^{\mu\nu} - 2 \eta \sigma^{\mu\nu}) - \frac{1}{2} \pi^{\mu\nu} \frac{\eta T}{\tau_\pi} d_\lambda \left(\frac{\tau_\pi}{\eta T} u^\lambda \right), \\
D\Pi &=& - \frac{1}{\tau_\Pi} (\Pi + \zeta \theta) - \frac{1}{2} \Pi \frac{\zeta T}{\tau_\Pi} d_\lambda \left(\frac{\tau_\Pi}{\zeta T} u^\lambda \right),
\end{eqnarray}
The equation of state (EoS) used in the simulations combines the lattice QCD based EoS from HotQCD collaboration~\cite{HotQCD:2014kol} at high temperatures with the hadron resonance gas EoS at low temperatures.

The fluid is converted to a particle ensemble at a constant temperature hypersurface (151 MeV) according to the Cooper-Frye prescription~\cite{Cooper:1974mv}, including shear and bulk viscous corrections. Then, the particles are fed to UrQMD~\cite{Bass:1998ca, Bleicher:1999xi} which describes the hadron rescattering and decays until kinetic freeze-out.


\section{Results}
\label{sec_result}
\subsection{Signatures of clustered nuclear configuration}
\begin{figure}
 \center
 \includegraphics[scale=0.97]{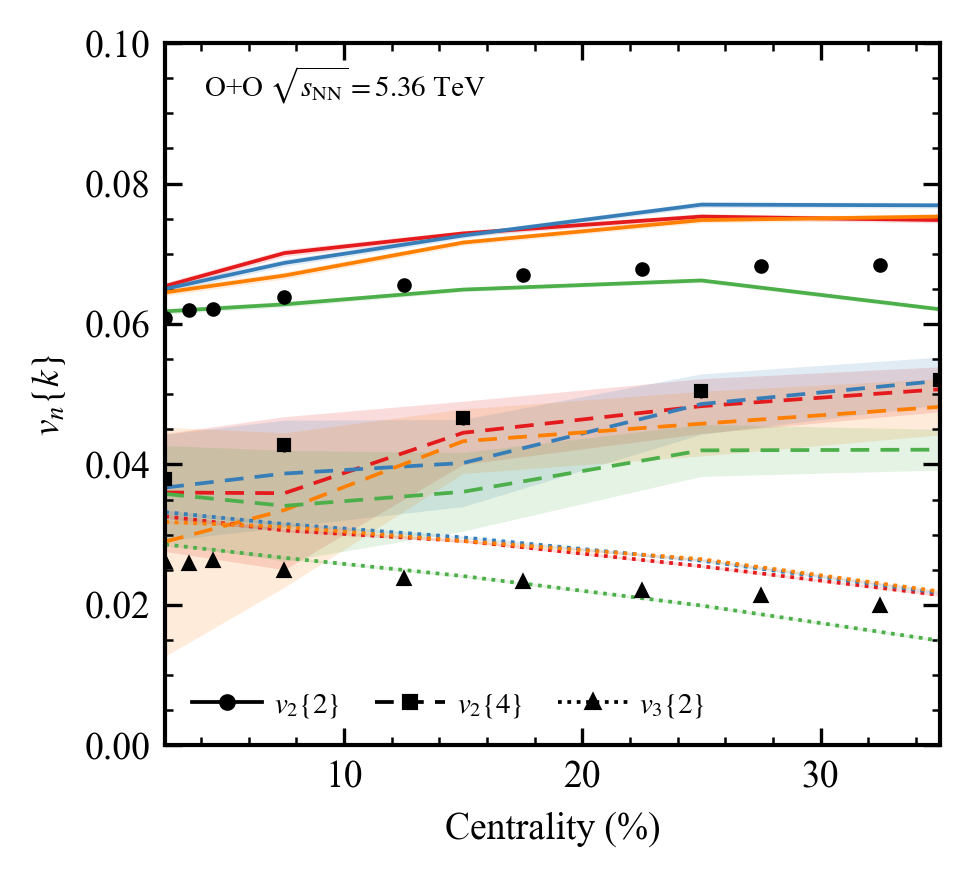}
 \includegraphics[scale=0.97]{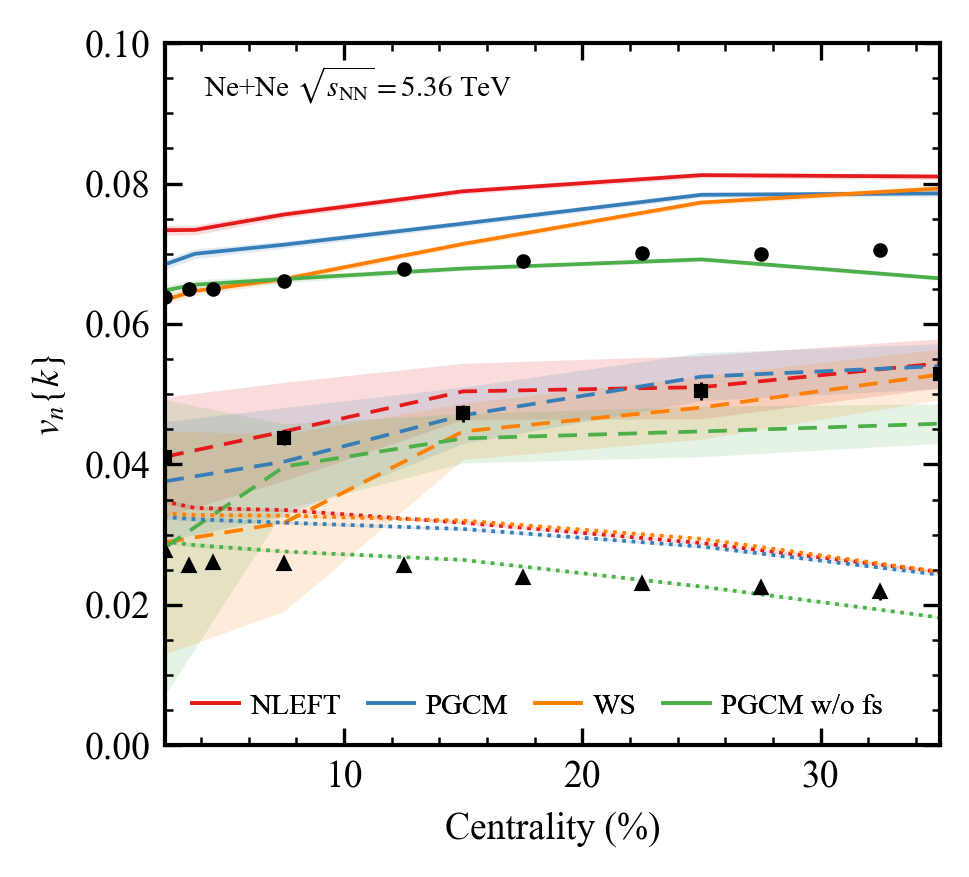}
  \caption{Flow coefficients $v_2\{2\}$, $v_2\{4\}$ and $v_3\{2\}$ are shown for O+O (left panel) and Ne+Ne (right panel) collisions with $|\eta| < 0.8$ and $0.2 < p_T < 3$ for three different initial profiles, and compared with ALICE~\cite{ALICE:2025luc} experimental data.}
  \label{fig:vn}
\end{figure}
The collective flow coefficients $v_2\{2\}$, $v_2\{4\}$ and $v_3\{2\}$ for O+O (left panel) and Ne+Ne (right panel) collisions, are plotted in Fig.~\ref{fig:vn} as a function of centrality. Fig.~\ref{fig:vn} shows that in the O+O collisions, there is little sensitivity of the flow observables to the presence of the clustered nuclear configurations in the initial state. An assumption of a round nucleus (Woods-Saxon) works equally well for the O+O system, but it clearly does not for Ne+Ne system. This difference arises since the predicted bowling-pin shaped deformation of ${}^{20}$Ne is more pronounced than the tetrahedral-shaped deformation of ${}^{16}$O, and therefore, $v_2\{2\}$ is strongly affected by the bowling-pin shape of ${}^{20}$Ne~\cite{Giacalone:2024luz}.
Here we note that $v_2\{2\}$ found with the pre-equlibrium free-streaming phase is slightly above the experimental data. The reason behind this is the MAP parameters found in the Bayesian analysis with Pb+Pb and p+Pb data in Ref.~\cite{Moreland:2018gsh} have been used in our study. However, it is argued in Ref.~\cite{Nijs:2020ors, Nijs:2021clz} that for p+Pb the posterior distributions are significantly
wider, and exclusion of the p+Pb data in the Bayesian analysis can lead to significant change of the MAP values of several parameters. Furthermore, inclusion of simulated O+O data in the Bayesian analysis in Ref.~\cite{Nijs:2021clz} decreases the uncertainties in the parameter values. Therefore, a global Bayesian analysis with real light ion collision experimental data is necessary to get more constrained MAP values of the parameters. We also note that, in our calculations, the 4-particle cumulant $c_2\{4\}$ becomes positive at peripheral collisions i.e., low multiplicity light-ion collisions, when the nucleon substructure is not included. The appearance of positive $c_2\{4\}$ i.e., the $c_2\{4\}$ sign puzzle, has been observed in pp collisions~ \cite{Zhao:2017rgg, Zhao:2020pty}. Therefore, determining the nucleon substructure width in the initial profiles from global Bayesian analysis with light-ion experimental data is important, since the flow in light-ion collisions is mainly  fluctuation driven~\cite{Krupczak:2026byr}.
\begin{figure}
 \center
 \includegraphics[scale=0.97]{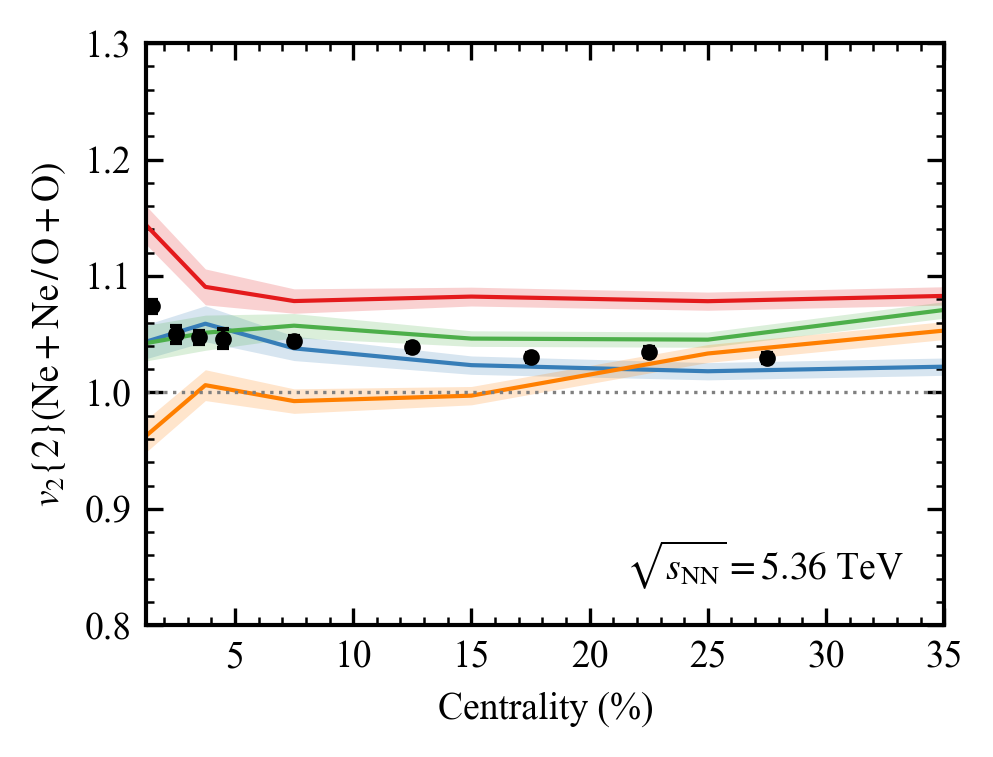}
 \includegraphics[scale=0.97]{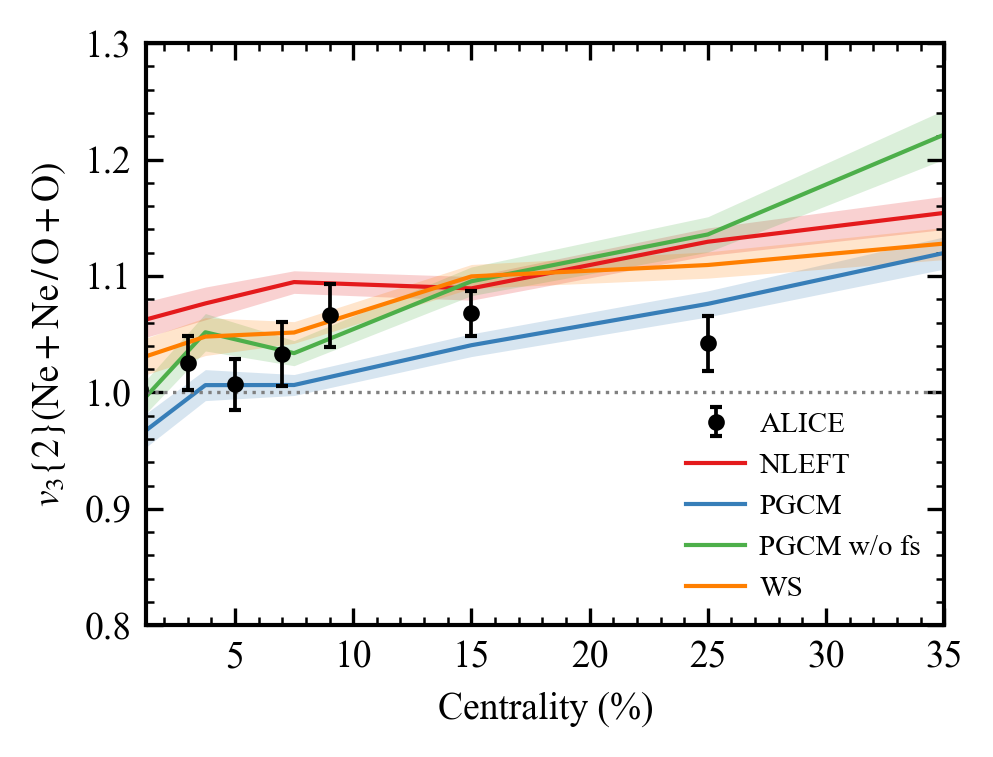}
  \caption{Flow coefficient ratios $v_2\{2\}$(Ne+Ne/O+O) (left panel) and $v_3\{2\}$(Ne+Ne/O+O) (right panel) with $|\eta| < 0.8$ and $0.2 < p_T < 3$  are plotted with centrality and compared with ALICE data~\cite{ALICE:2025luc}.}
  \label{v22_v32_NebyO}
\end{figure}

\begin{figure}
 \center
 \includegraphics[scale=0.93]{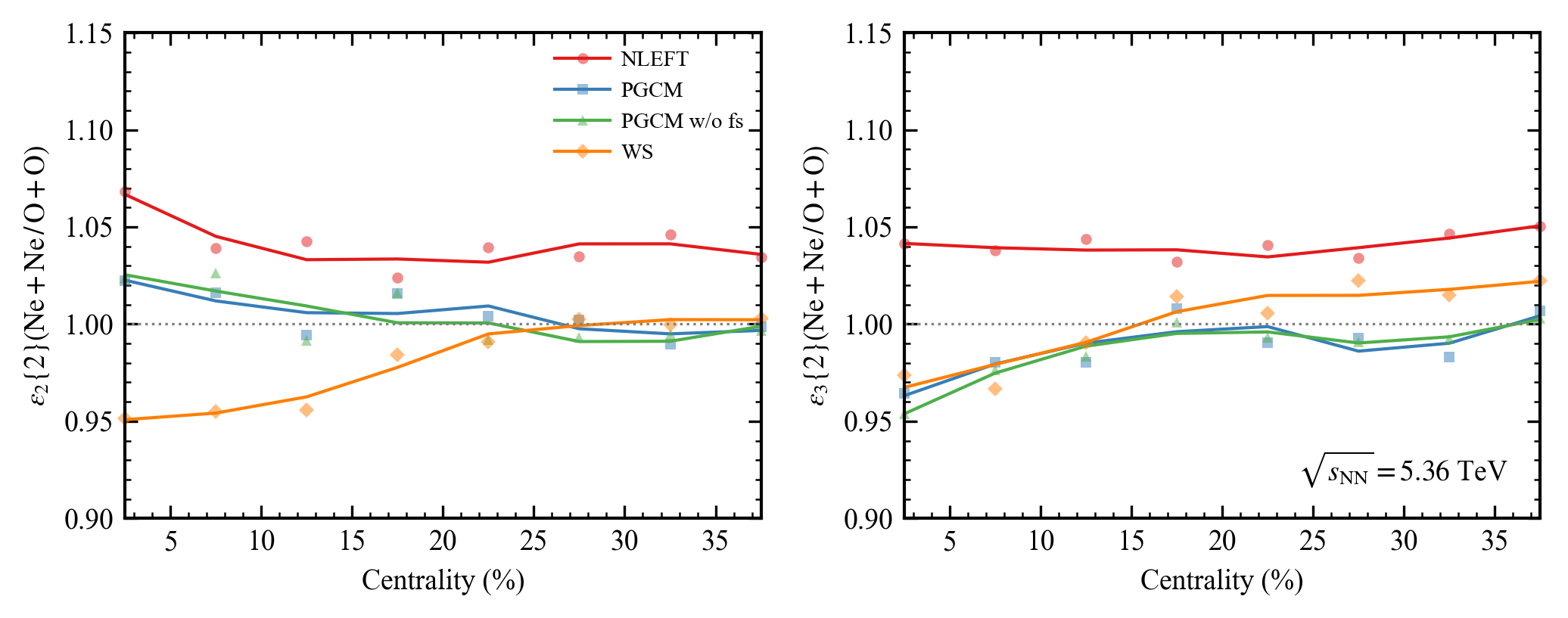}
 \caption{The eccentricity ratio of the initial entropy density profiles $\varepsilon_2\{2\}$(Ne+Ne/O+O) (left panel) and $\varepsilon_3\{2\}$(Ne+Ne/O+O) (right panel) (with Savitzky-Golay smoothing) are plotted with centrality for different nuclear configurations.}
  \label{eps_ratio}
\end{figure}
\begin{figure}
 \center
 \includegraphics[scale=0.93]{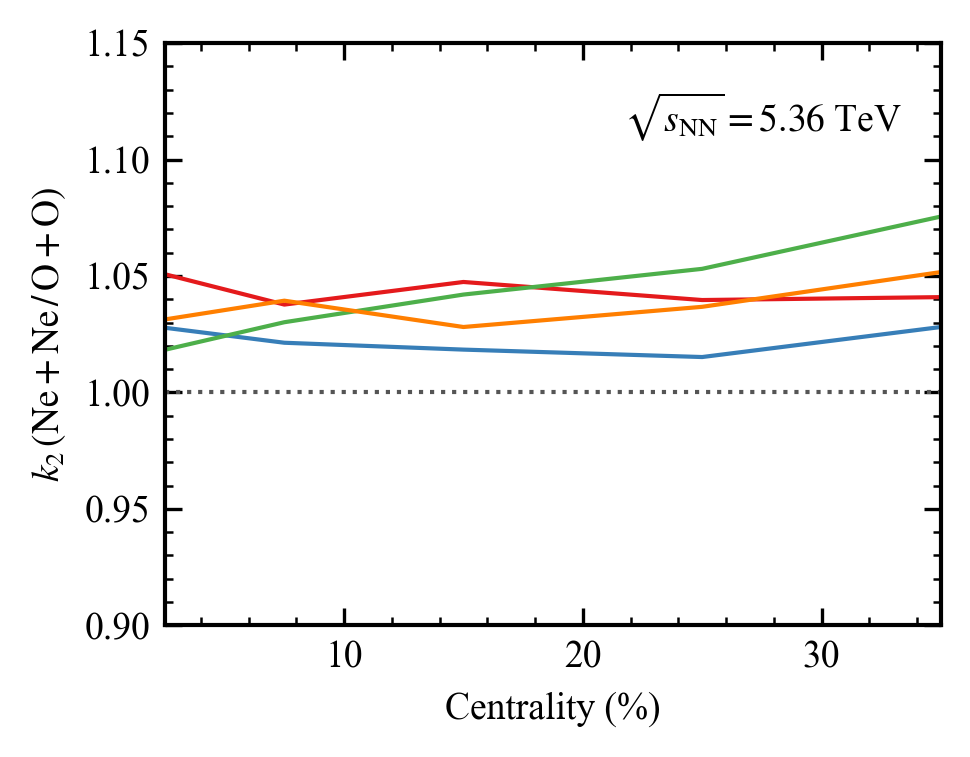}
 \includegraphics[scale=0.93]{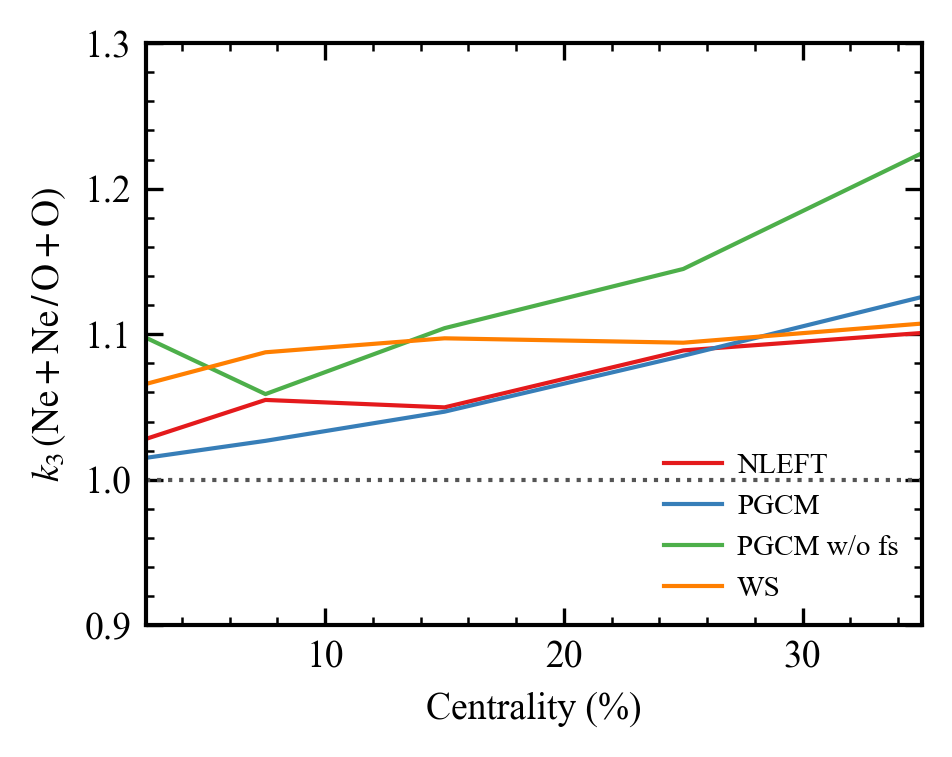}
 \caption{The centrality dependence of the medium response ratio $k_2$\,(Ne+Ne/O+O) and $k_3$\,(Ne+Ne/O+O) corresponding to Fig.~\ref{v22_v32_NebyO} and \ref{eps_ratio} is shown.}\label{response_ratio}
\end{figure}

\begin{figure}
 \center
 \includegraphics[scale=0.97]{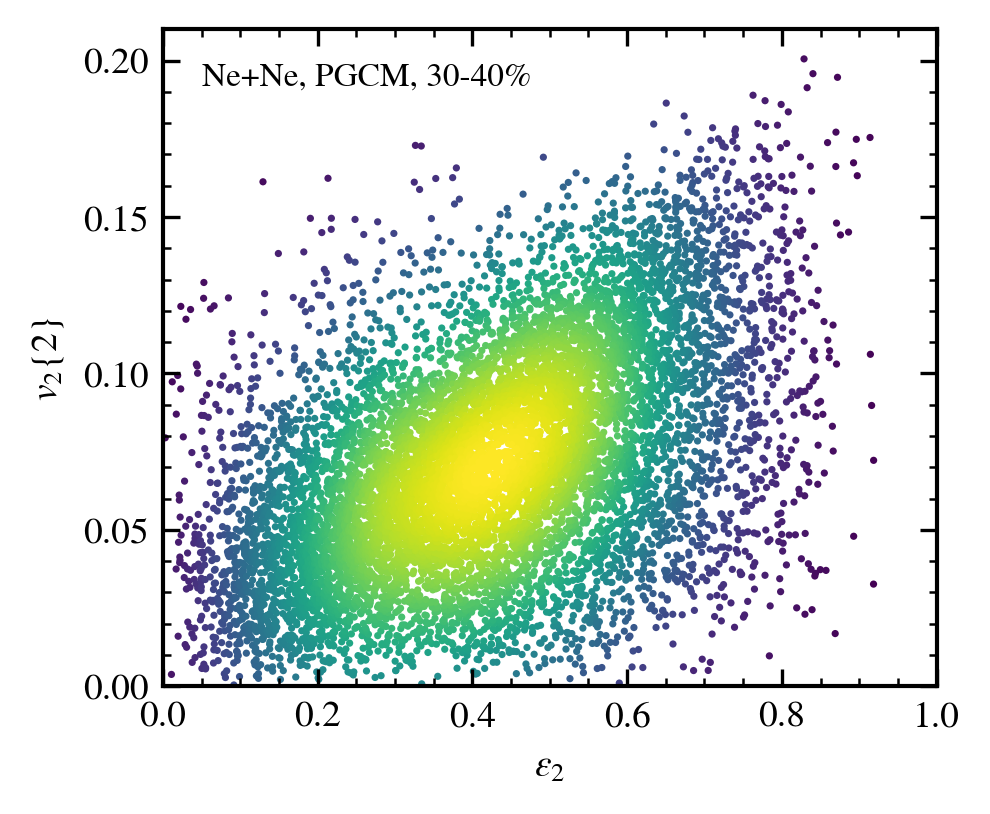}
 \includegraphics[scale=0.975]{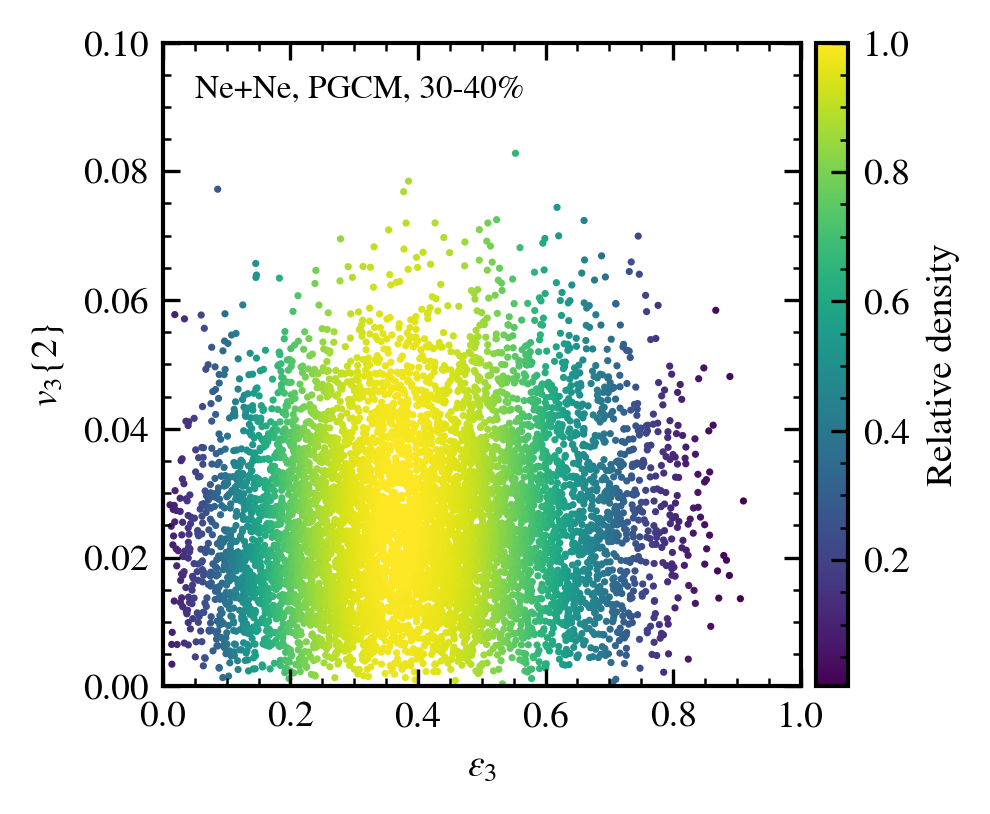}
 \caption{$\varepsilon_2$ and $v_2\{2\}$ (left panel), and $\varepsilon_3$ and $v_3\{2\}$ (right panel) are shown for Ne+Ne collisions with PGCM initial profiles at 30-40\% centrality class.}
 \label{flow_eccentricity_corr}
\end{figure}

The left and right panel of Fig.~\ref{v22_v32_NebyO} show the centrality dependence of $v_2\{2\}$(Ne+Ne)$/$ $v_2\{2\}$(O+O) and $v_3\{2\}$(Ne+Ne)$/$ $v_3\{2\}$(O+O), respectively. For both clustered nuclear configurations $v_2\{2\}$(Ne+Ne/O+O) $>$ 1, and it peaks in the most central events. WS nuclear configurations show the opposite trend, which we explain below. The centrality dependence of the initial eccentricity ratios $\varepsilon_2\{2\}$(Ne+Ne/O+O) and $\varepsilon_3\{2\}$(Ne+Ne/O+O) is shown for different nuclear configurations in Fig.~\ref{eps_ratio}. The event-averaged response ratio $k_2$(Ne+Ne/O+O), where $k_2 \equiv v_2/\varepsilon_2$, is slightly above 1 and mostly flat with centrality, as shown in the left panel of Fig.~\ref{response_ratio}. That means for $v_2$, the medium response almost cancels because the medium properties (e.g., $\eta/s$) are similar and the lifetime of the collision is almost the same for Ne+Ne and O+O collisions. Therefore, $v_2\{2\}$(Ne+Ne)$/$ $v_2\{2\}$(O+O) effectively represent the initial nuclear geometry. As mentioned earlier, $v_2\{2\}$ is strongly affected by the bowling-pin shape of ${}^{20}$Ne. Owing to its ${}^{16}$O+$\alpha$ cluster structure, ${}^{20}$Ne carries a large intrinsic quadrupole deformation, together with an octupole component~\cite{Giacalone:2024luz}. In contrast, the tetrahedral $\alpha$-cluster configuration of ${}^{16}$O is dominated by an octupole deformation, with only a small quadrupole deformation. Since the mean-square eccentricity in central collisions follows $\varepsilon_2^2 \approx a_2' + b_2' \beta_2^2$~\cite{Jia:2021qyu}, where $\beta_2$ characterizes the overall quadrupole deformation, the quadrupole deformation of ${}^{20}$Ne enhances $\varepsilon_2$ and, consequently, $v_2$ relative to O+O collisions. This is reflected in the ratio $v_2\{2\}$(Ne+Ne)/$v_2\{2\}$(O+O), which exceeds unity and is largest in the
most central collisions, consistent with the ALICE measurement~\cite{ALICE:2025luc}. Towards peripheral collisions, $\varepsilon_2$ becomes dominated by the fluctuations of the overlap region geometry, and the contribution of the intrinsic nuclear shape becomes subleading. On the other hand, for WS configurations,  $v_2\{2\}$(Ne+Ne/O+O) $<$ 1 at central collisions since both ${}^{20}$Ne and ${}^{16}$O have a spherical shape, and ${}^{20}$Ne, due to its bigger size, has less fluctuation compared to ${}^{16}$O, which leads to lower $\varepsilon_2$ for ${}^{20}$Ne compared to ${}^{16}$O. This can be clearly seen in the left panel of Fig.~\ref{eps_ratio}. On the other hand, $v_3\{2\}$ is sensitive to the octupolar deformation of the colliding nuclei, on
top of the dominant contribution from nucleon-position fluctuations~\cite{Giacalone:2024luz}. For the clustered NLEFT and PGCM configurations, both ${}^{20}$Ne and ${}^{16}$O carry
octupole deformation, so that their contributions to $\varepsilon_3$ largely cancel in the ratio and $v_3\{2\}$(Ne+Ne/O+O) remains close to unity in the most central collisions. The centrality dependence of $\varepsilon_3\{2\}$(Ne+Ne/O+O) is shown in the right panel of Fig.~\ref{eps_ratio}. The event-averaged response ratio $k_3$(Ne+Ne/O+O), where
$k_3 \equiv v_3/\varepsilon_3$, exhibits a slightly increasing trend towards peripheral collisions, as shown in the right panel of Fig.~\ref{response_ratio}. Therefore, the medium effects do not fully cancel in the ratio $v_3\{2\}$(Ne+Ne)$/$ $v_3\{2\}$(O+O), particularly in peripheral collisions. This can be attributed to the absence of an apparent correlation between $v_3\{2\}$ and $\varepsilon_3$ in peripheral collisions, as illustrated in the right panel of Fig.~\ref{flow_eccentricity_corr}. This behaviour contrasts with the much stronger correlation observed between $v_2\{2\}$ and $\varepsilon_2$, shown in the left panel of Fig.~\ref{flow_eccentricity_corr}. We also present the results obtained for the case with no pre-equlibrium free-streaming. The impact of pre-equilibrium free-streaming on the observables is discussed in the following subsection.

\begin{figure}
 \center
 \includegraphics[scale=1]{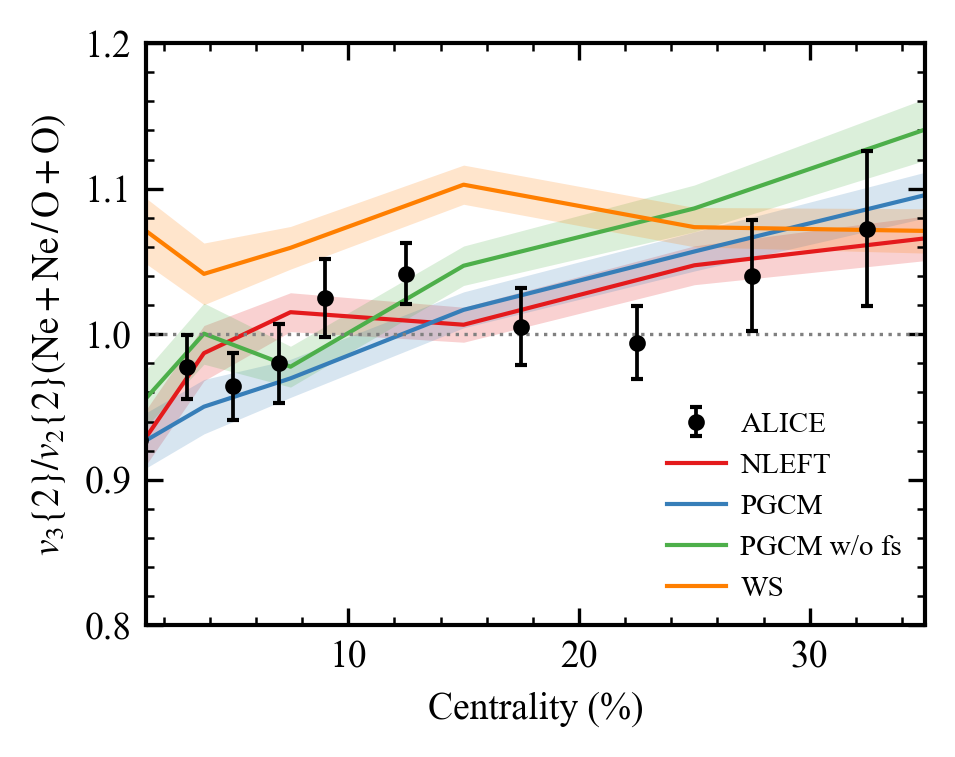}
  \caption{  $v_3\{2\}/v_2\{2\}$(Ne+Ne/O+O) with centrality is presented for three different nuclear configuration with $|\eta| < 0.8$ and $0.2 < p_T < 3$, and compared with ALICE data~\cite{ALICE:2025luc}.}
  \label{v32overv22}
\end{figure}

Fig.~\ref{v32overv22} shows the centrality dependence of the double ratio $[v_3\{2\}/v_2\{2\}]_{\rm Ne+Ne}/[v_3\{2\}/v_2\{2\}]_{\rm O+O}$ for the different initial configurations. In central collisions, the clustered NLEFT and PGCM configurations yield clearly lower values than the WS baseline. Since both ${}^{20}$Ne and ${}^{16}$O carry octupole deformation, their contributions to
$v_3\{2\}$ largely cancel in the ratio, and the suppression of the double ratio is driven mainly by the enhancement of $v_2\{2\}$ in Ne+Ne collisions. It therefore reflects the large quadrupole deformation of ${}^{20}$Ne.

\begin{figure}
 \center
 \includegraphics[scale=0.97]{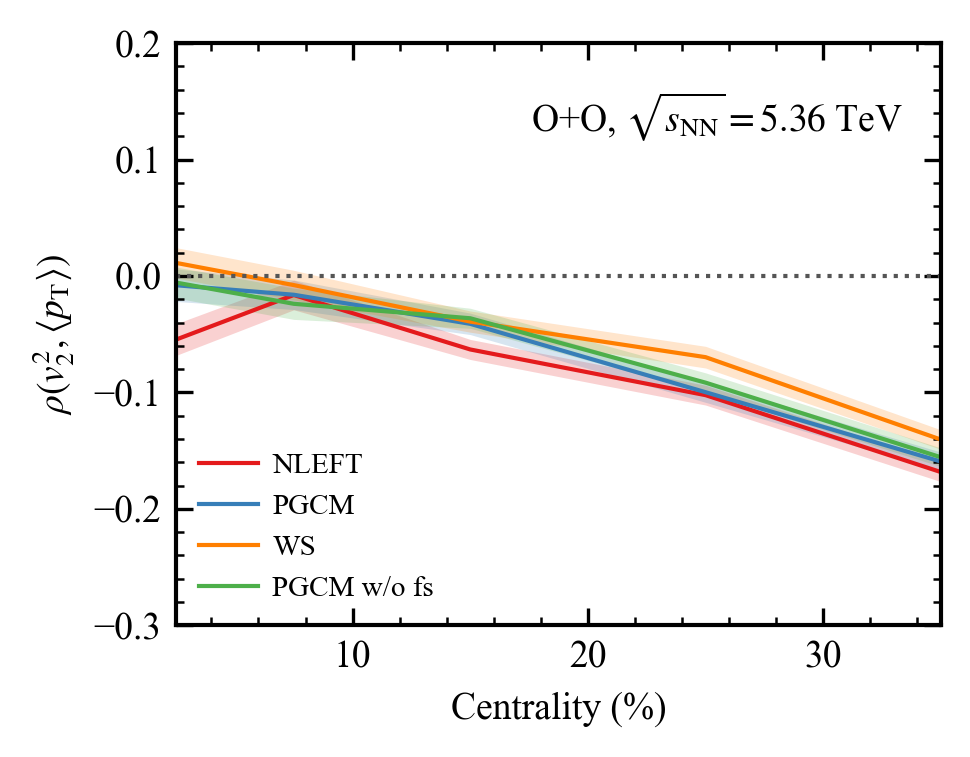}
  \includegraphics[scale=0.97]{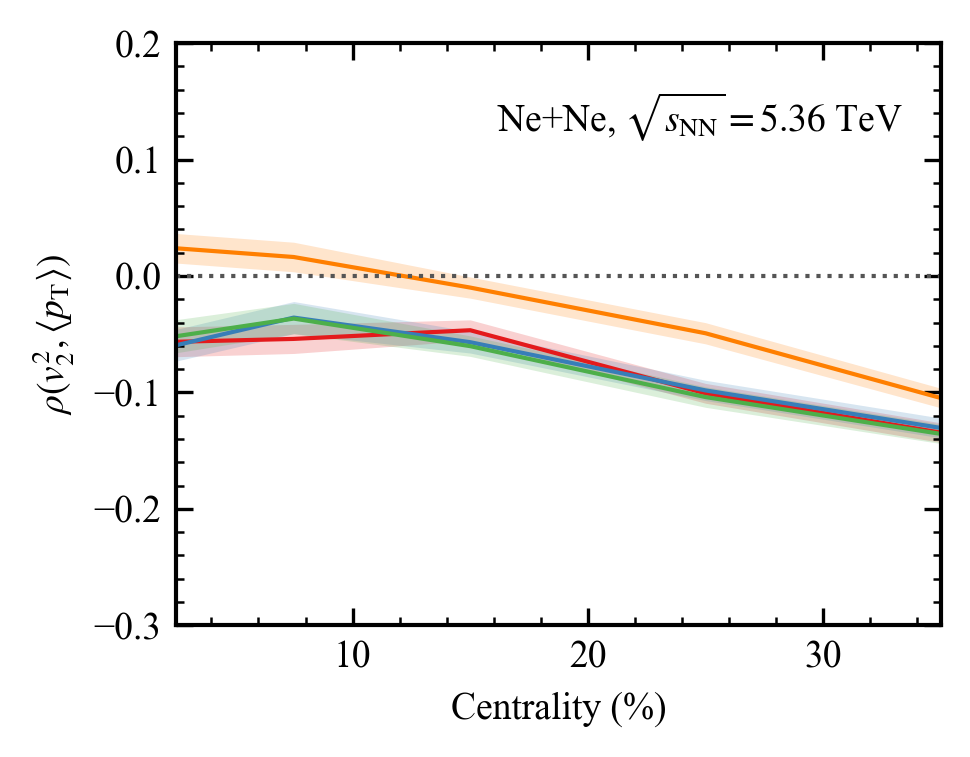}
  \caption{The centrality dependence of the Pearson correlation coefficient $\rho(v_2\{2\}^2, \langle p_T \rangle)$ with $|\eta| < 0.8$ and $0.2 < p_T < 3$ for O+O (left panel) and Ne+Ne (right panel) collisions is shown.}
  \label{pearson_OO_NeNe}
\end{figure}
\begin{figure}
 \center
 \includegraphics[scale=0.97]{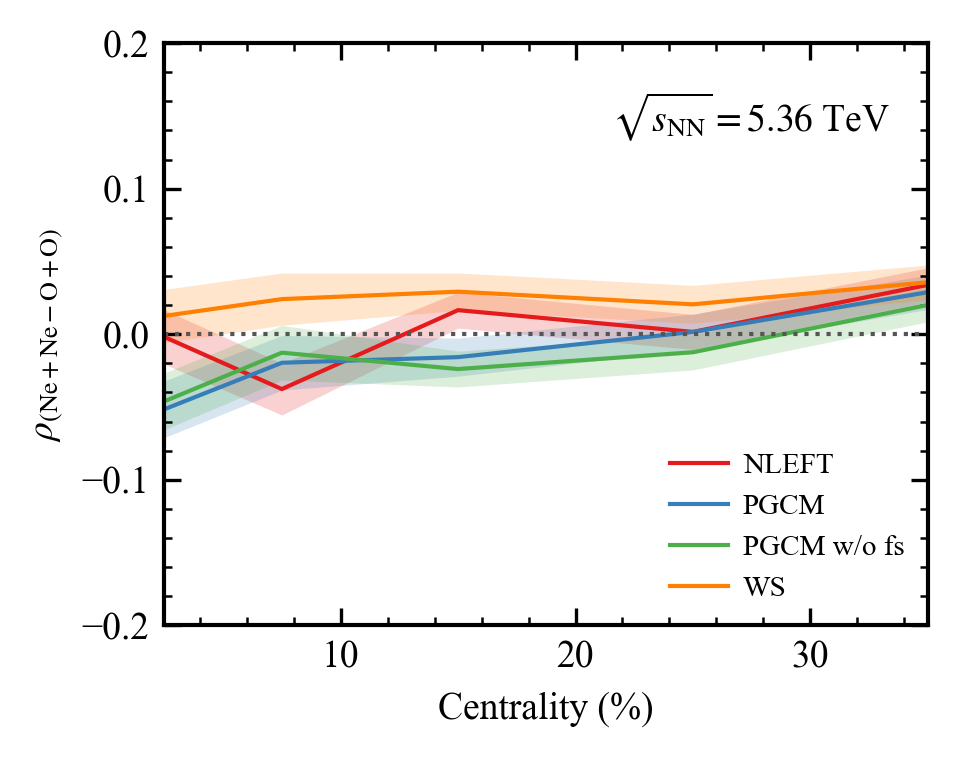}
  \caption{$\rho_{(\rm Ne+Ne - \rm O+O)}$ for different initial nuclear structures are presented.}
  \label{pearson_diff}
\end{figure}
The Pearson coefficient $\rho_2\equiv\rho(v_2\{2\}^2, \langle p_T \rangle)$~\cite{Bozek:2016yoj} is a sensitive probe of the deformation of the colliding nuclei since it represents the correlation of the shape and size of the created fireball~\cite{Giacalone:2024luz, Bozek:2020drh}. For O+O collisions, $\rho_2$ is close to zero in central collisions and becomes negative towards peripheral collisions as shown in the left panel of Fig.~\ref{pearson_OO_NeNe}, with negligible dependence on the nuclear structure input. Since $\rho_2$ is sensitive  to  $\beta_2^3$~\cite{Jia:2021qyu, Giacalone:2024luz}, it requires a well-defined prolate or oblate shape to show significant anticorrelation between the elliptic flow coefficient and $\langle p_T \rangle$ in the central events. The tetrahedral $\alpha$-cluster configuration of ${}^{16}$O has no static intrinsic quadrupole moment~\cite{Dudek:2002mq,Bijker:2014tka}, so that its effective quadrupole
deformation is small~\cite{Giacalone:2024luz,Pritychenko:2013gwa} and its contribution to $\rho_2$ is strongly suppressed. For Ne+Ne collisions shown in the right panel of Fig.~\ref{pearson_OO_NeNe}, the clustered NLEFT and PGCM inputs yield negative values already in central collisions, whereas the WS profile gives a positive correlation. This suppression reflects the prolate bowling-pin shape of ${}^{20}$Ne. In central collisions, configurations with large ellipticity are associated with larger overlap areas, inducing an anticorrelation between $v_2\{2\}^2$ and $\langle p_T \rangle$. The difference $\rho_2$(Ne+Ne)$\,-\,\rho_2$(O+O) shown in Fig.~\ref{pearson_diff} isolates this effect; while the WS baseline remains positive at all centralities, the clustered inputs give a negative difference in central collisions, which approaches the baseline towards peripheral collisions as the fluctuation of the overlap area dominates over the intrinsic nuclear shape.
\subsection{Effect of pre-equilibrium free streaming}

\begin{figure}
 \center
 \includegraphics[scale=0.97]{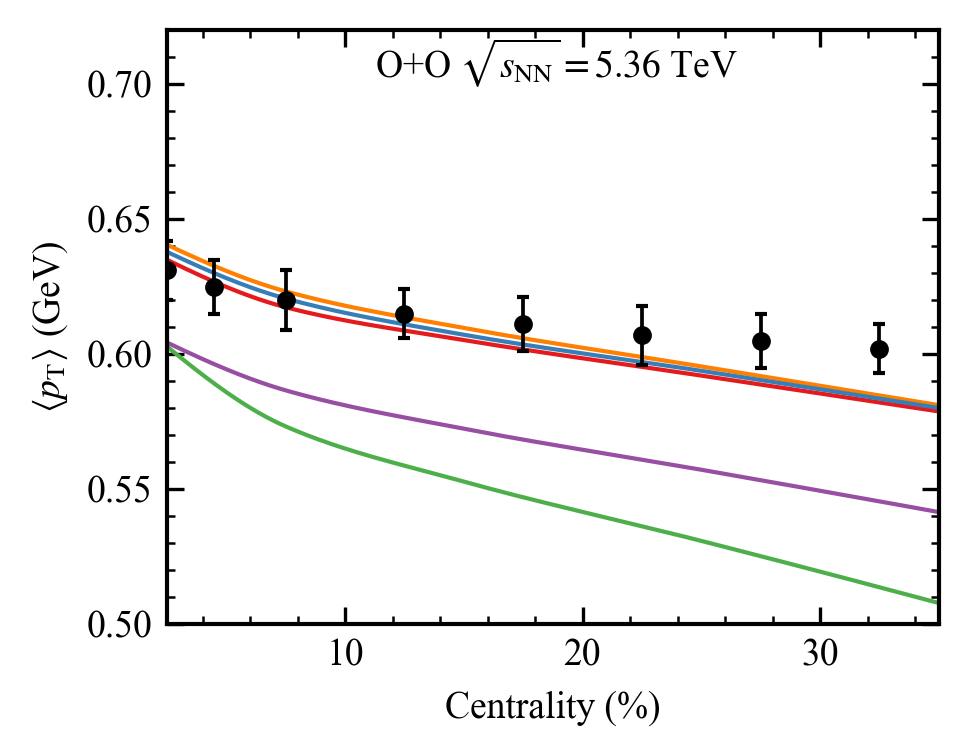}
  \includegraphics[scale=0.97]{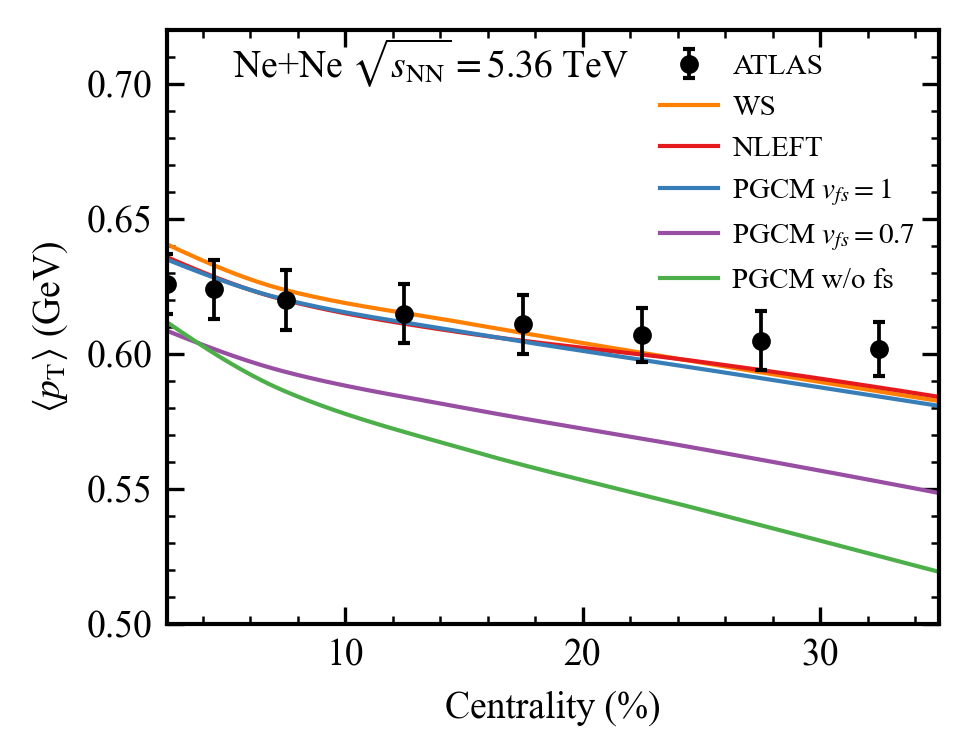}
  \caption{The centrality dependence of $\langle p_T \rangle$ at midrapidity ($|\eta|<0.5$) is shown for different nuclear configurations and compared with ATLAS~\cite{ATLAS:2026zgq} experimental data.}
\label{meanpt_OO_NeNe}
\end{figure}
The mean transverse momentum of the charged particles with $p_T>0$ and $|\eta|<0.5$ for different nuclear configurations with centrality are shown in Fig.~\ref{meanpt_OO_NeNe}.
To quantify the role of the pre-equilibrium stage, we compare three scenarios: free streaming with $v_{fs}$ = 1, free streaming with reduced velocity $v_{fs}$ = 0.7, similar to what has been used in Trajectum calculations, and no free streaming, re-tuning the T$_{\text{R}}$ENTo normalization in each case to reproduce the measured charged-particle multiplicity. While $\langle p_T \rangle$ is insensitive to the nuclear structure of ${}^{16}$O and ${}^{20}$Ne, it varies significantly between these scenarios. Because the pre-equilibrium stage occupies a substantial fraction of the short lifetime of light-ion systems, the transverse flow developed before fluidisation contributes a considerably larger share of the final $\langle p_T \rangle$ than in large systems. The $\langle p_T \rangle$ in O+O and Ne+Ne collisions therefore provides a sensitive constraint on pre-equilibrium dynamics.

\begin{figure}
 \center
 \includegraphics[scale=0.97]{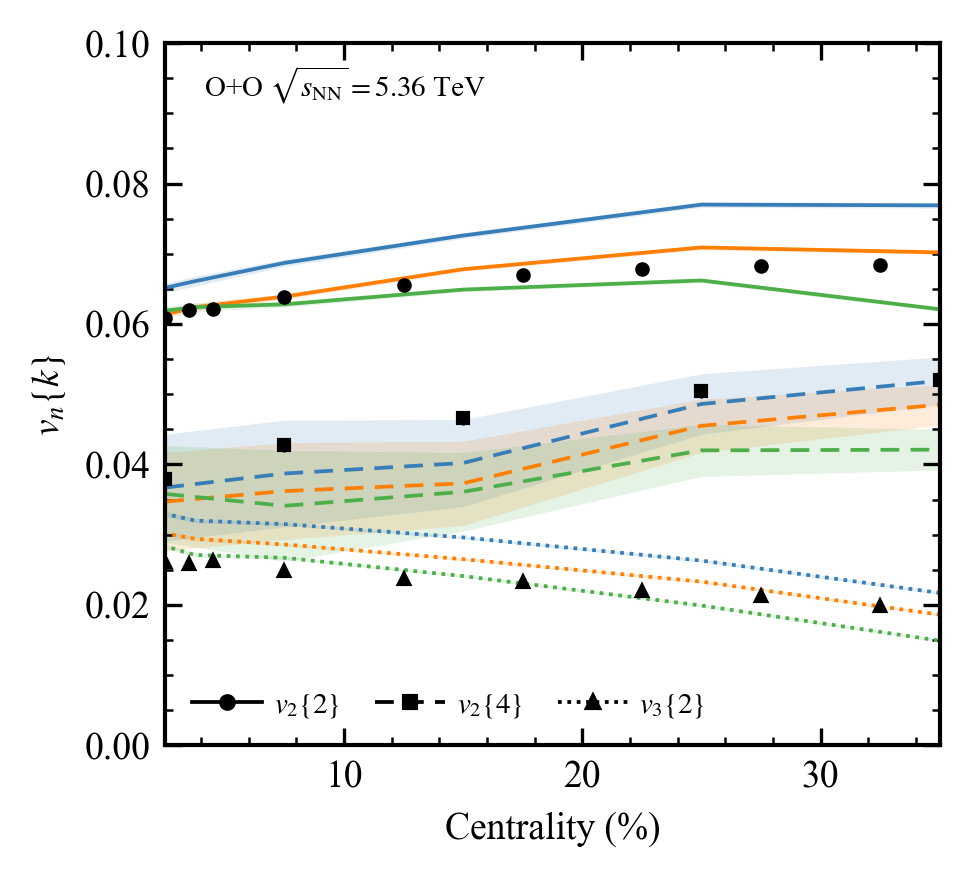}
  \includegraphics[scale=0.97]{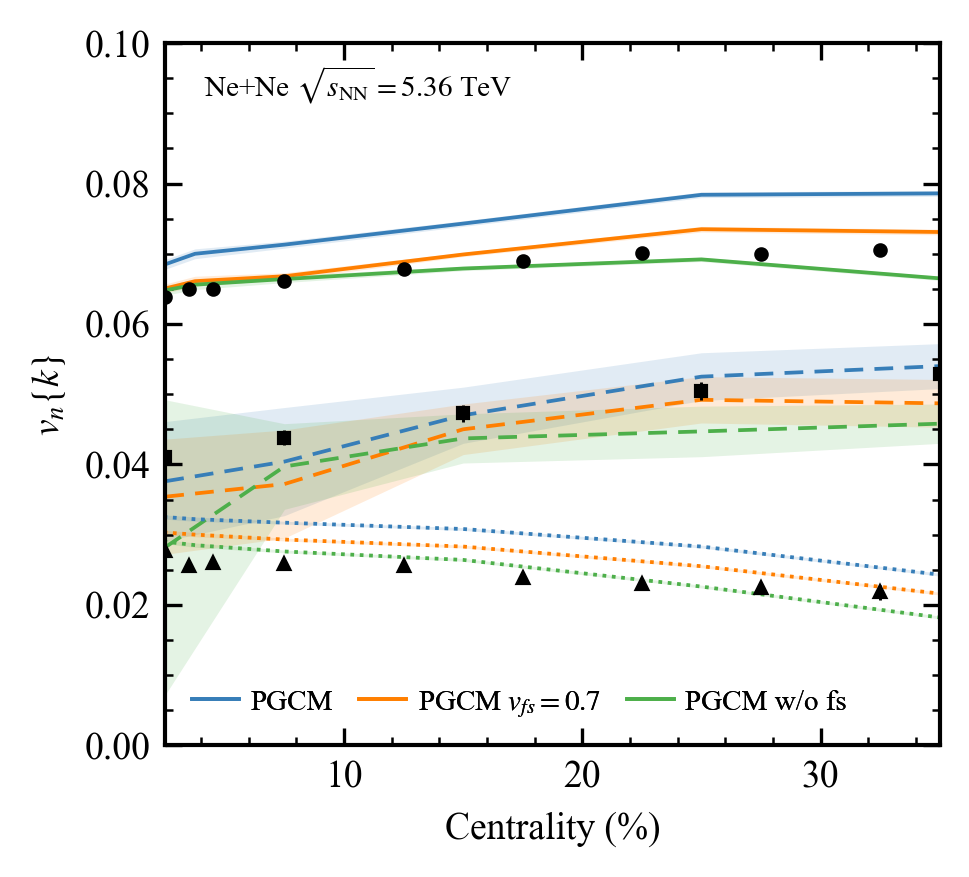}
  \caption{The effect of pre-equlibrium stage with different free-streaming velocities on the flow coefficients with $|\eta| < 0.8$ and $0.2 < p_T < 3$ for O+O (left panel) and Ne+Ne (right panel) systems is presented and compared with ALICE data~\cite{ALICE:2025luc}.}
\label{vn_OO_NeNe_fs}
\end{figure}
To assess the role of the pre-equilibrium stage, we compare the flow coefficients calculated in three scenarios: free streaming with $v_{fs}$ = 1, free streaming with reduced velocity $v_{fs}$ = 0.7, and no free streaming, with hydrodynamics initialized at $\tau=0.48$ fm for O+O (left panel) and Ne+Ne (right panel) collisions in Fig.~\ref{vn_OO_NeNe_fs}. The pre-equilibrium stage significantly affects the anisotropic flow.  $v_2\{2\}$, $v_2\{4\}$ and $v_3\{2\}$ are largest for free streaming with $v_{fs}$ = 1, reduced for $v_{fs}$ = 0.7, and smallest without free streaming. The initial eccentricity $\varepsilon_2$ changes only slightly during free streaming (by less than 5\% across centrality). The differences in the flow coefficients therefore originate from the collective flow and anisotropic stress generated during the pre-equilibrium stage, which provide a non-zero momentum anisotropy at the onset of hydrodynamics. Together with the behaviour of $\langle p_T \rangle$ discussed above, these results show that in light-ion collisions a substantial fraction of both radial and anisotropic flow develops before hydrodynamization. Remarkably, despite this strong sensitivity of the individual flow coefficients, the ratio $v_2\{2\}$(Ne+Ne/O+O)  is almost unaffected by the choice of pre-equilibrium scenario due to the similar sizes and lifetimes of the two systems. On the other hand, $v_3\{2\}$(Ne+Ne/O+O) exhibits a mild sensitivity, particularly in peripheral collisions, as shown in Fig.~\ref{v22_v32_NebyO}, apparently due to the lack of correlation between $v_3\{2\}$ and $\varepsilon_3$. The difference $\rho_2$(Ne+Ne)$\,-\,\rho_2$(O+O) is largely insensitive to the pre-equilibrium free-streaming phase: the results obtained with and without free streaming overlap within statistical uncertainties, as shown in
Fig.~\ref{pearson_diff}. This demonstrates that the flow ratio $v_2\{2\}$(Ne+Ne/O+O) and the difference $\rho_2$(Ne+Ne)$\,-\,\rho_2$(O+O) are reasonably stable against uncertainties in the pre-equilibrium dynamics, which strengthens their use as probes of the nuclear structure of ${}^{16}$O and ${}^{20}$Ne. In contrast, the mild sensitivity of the ratio $v_3\{2\}$(Ne+Ne/O+O) to the free-streaming phase suggests that it may not serve as a robust probe of nuclear structure.

\section{Summary}
\label{sec_summary}
We have studied O+O and Ne+Ne collisions at $\sqrt{s_{\rm NN}} = 5.36$~TeV within a hybrid framework of T$_{\text{R}}$ENTo+VISHNew+UrQMD, using Woods-Saxon, NLEFT and PGCM nuclear configurations for ${}^{16}$O and ${}^{20}$Ne. The differences between the nuclear configurations
appear more clearly in the flow coefficients of Ne+Ne collisions than in those of O+O, since the large quadrupole deformation of ${}^{20}$Ne strongly modifies the initial geometry, whereas in O+O the structure effects remain small compared to the fluctuation-driven anisotropies. The ratio $v_2\{2\}$(Ne+Ne/O+O) is enhanced in
central collisions for the clustered configurations, reflecting the bowling-pin shape of ${}^{20}$Ne. For $v_3\{2\}$(Ne+Ne/O+O), the octupole contributions of both nuclei largely cancel, keeping the ratio close to unity for the clustered configurations in the most central events. However, since $v_3\{2\}$ is largely affected by fluctuations, which are enhanced in light-ion systems, it is difficult to draw firm conclusions about its dependence on the initial nuclear configurations. The double ratio $[v_3\{2\}/v_2\{2\}]_{\rm Ne+Ne}/[v_3\{2\}/v_2\{2\}]_{\rm O+O}$ is suppressed in central collisions for clustered configurations, driven mainly by the quadrupole deformation of ${}^{20}$Ne. The Pearson coefficient $\rho_2$ is insensitive to the structure of ${}^{16}$O, whereas the prolate shape of ${}^{20}$Ne yields negative values in central Ne+Ne collisions; the difference $\rho_2$(Ne+Ne)$\,-\,\rho_2$(O+O) is thus
negative in central collisions for the clustered configurations and converges onto the Woods-Saxon baseline in peripheral collisions. Finally, $\langle p_T \rangle$ and the individual flow coefficients depend strongly on the pre-equilibrium stage, since a substantial fraction of the radial and anisotropic flow develops before hydrodynamization in these short-lived systems, while the
initial eccentricities change only slightly. We find tension between the description of flow observables and $\langle p_T \rangle$ in both O+O and Ne+Ne collisions. The former is consistent with ``mild'' pre-hydro free-streaming as in \textsc{Trajectum} calculations, whereas the latter necessitates a strong free-streaming. In contrast, the $\rho_2$ difference is essentially independent of the pre-equilibrium scenario, establishing it as a robust probe of nuclear structure in light-ion collisions.
\section{Acknowledgement}
BK and IK were supported by the Czech Science Foundation, under No 25-16877S. Computational resources were provided by the e-INFRA CZ project (ID:90254), supported by the Ministry of Education, Youth and Sports of the Czech Republic.
\bibliography{draft}
\end{document}